\documentstyle[subequation,epsfig,aps]{revtex}
\newcommand{\dl}{L_W}
\newcommand{\tvet}{\mbox{\boldmath $t$}}
\newcommand{\Ttot}{\mbox{\boldmath T}_{{\rm t}}}
\newcommand{\ov}{{\cal O}}

\newcommand  \f  \varphi
\newcommand \bra {\langle}
\newcommand \ket {\rangle}
\newcommand{\be}{\begin{equation}}
\newcommand{\ee}{\end{equation}}
\newcommand{\ben}{\begin{displaymath}}
\newcommand{\een}{\end{displaymath}}
\newcommand{\ba}{\begin{eqnarray}}
\newcommand{\ea}{\end{eqnarray}}
\newcommand{\ban}{\begin{eqnarray*}}
\newcommand{\ean}{\end{eqnarray*}}
\newcommand{\cro}{\dagger}

\newlength{\www}

\newcommand{\kvet}{\mbox{\boldmath $k$}}

\begin{document}


\title{\Large\bf Electroweak double logarithms in inclusive
observables for a generic initial state}
\author{\large Marcello Ciafaloni}
\address{\it Dipartimento di Fisica, Universit\`a di Firenze e
\\ INFN - Sezione di Firenze,
 I-50125 Florence , Italy\\
E-mail: ciafaloni@fi.infn.it}
\author{\large Paolo Ciafaloni}
\address{\it INFN - Sezione di Lecce,
\\Via per Arnesano, I-73100 Lecce, Italy
\\ E-mail: paolo.ciafaloni@le.infn.it}
\author{\large Denis Comelli}
\address{\it INFN - Sezione di Ferrara,
\\Via Paradiso 12, I-35131 Ferrara, Italy\\
E-mail: comelli@fe.infn.it}
\maketitle
\vskip0.3cm
\begin{abstract}
High energy observables 
are characterized by large  electroweak radiative
corrections of infrared origin;  double logarithms  are present
even for inclusive cross sections, thus violating the
Bloch Nordsieck theorem. This effect,
related to the initial states carrying nonabelian isospin charges, 
is here investigated for {\sl any} inclusive cross section
in the SU(2)$\otimes$U(1) symmetric limit, that is appropriate for
energies much higher than the weak scale.
We develop a general formalism allowing to calculate the 
all order double log resummed cross sections once the hard (tree level)
ones are known. The relevant cases of fermion-fermion, fermion-boson
and boson-boson scattering are discussed.
\end{abstract}


\def\baselinestretch{1.1}
\section{Introduction}
The behaviour of electroweak (EW) cross sections for energies $\sqrt{s}$ much  
larger than the electroweak scale $M\sim$ 100 GeV is a theoretically
interesting (and difficult) problem and a phenomenologically relevant
one. The discovery of the physical origin of the leading asymptotic
behaviour, related to the infrared (IR) structure of the electroweak theory
\cite{cc1}, has opened the way to various attempt of all orders
resummation, leading 
sometimes to controversial results, and to 
surprising features in some other cases. 
 On one hand,
electroweak symmetry breaking makes the calculation of exclusive
quantities (i.e., observables including only photon emission)
particularly hard, so that different results are present in
the literature \cite{cc2}. On the other hand, precisely because of the
unique features of weak interactions with respect to strong ones for
instance, unsuppressed double logs (DL) of infrared 
origin\footnote{these (double) logarithms $\sim \log^2\frac{\sqrt{s}}{M}$
occur because at energies much larger than the
EW scale $M_Z\approx M_W\equiv M$, the latter acts as a cutoff for the
collinear and IR divergences that would be present in the vanishing
$M$ limit} are present also 
in inclusive quantities 
(where also $W,Z$ emission is included),
leading to  violation of the Bloch-Nordsieck theorem \cite{3p1}. 

In this paper we develop a general formalism that allows to obtain
the expression for the leading EW radiative corrections to inclusive
cross sections  
for any given hard process. 
Through this formalism, once the hard (Born level) cross section is
known, resummation of EW effects  at all orders at the leading (double
log) level
is easily obtained.

We work in the SU(2)$\otimes$U(1) symmetric limit,
with all invariants much bigger than the Higgs  
and heavy gauge bosons masses $M_W\sim M_Z\sim M_H\equiv M$.
To be specific, we consider processes with two partons (e.g. leptons,
quarks, or bosons) in the initial
state, 
characterized by a single hard scale, typically
the c.m energy $\sqrt{s}$, much greater than 
the EW symmetry breaking scale
$M$, and including emission of soft 
 weak bosons $\gamma,Z,W$ of energy $\omega\ll \sqrt{s}$.
All other hard scales are of the same order,
namely $|s|\sim |t|\sim |u|\gg M$.
As has been noticed in \cite{3p1},
despite being inclusive, 
this kind of process is characterized 
by large double logs $\sim \log^2\frac{\sqrt{s}}{M}$
of infrared origin, due to the fact that initial state particles
carry nonabelian charges (weak isospin) which are {\it fixed}
by the accelerator; this in contrast
to QCD for instance where confinement forces averaging over initial
colour \cite{dft}.

Unlike the exclusive case, the radiative corrections to inclusive
observables considered here are sensitive to the weak scale $M$ only,
because this scale represents the threshold for both the 
nonabelian double logs
and gauge symmetry restoration at the same time. For this reason, 
exponentiation to all orders was easily proved in \cite{3p}, along the 
same lines as QCD, with a single IR cutoff $M$;
this point is rediscussed in next Section. 
While only fermions were considered in the initial state
in \cite{3p}, we extend here the analysis to {\it any} initial particles
belonging to a weak isospin multiplet, including gauge bosons.
This generalization is essential in order to treat boson fusion processes,
which are important for hadronic accelerators, like LHC.

\section{Resummed overlap matrix}
\begin{figure}[htb]\setlength{\unitlength}{1cm}
\begin{picture}(12,7.5)
\put(0.3,1){\epsfig{file=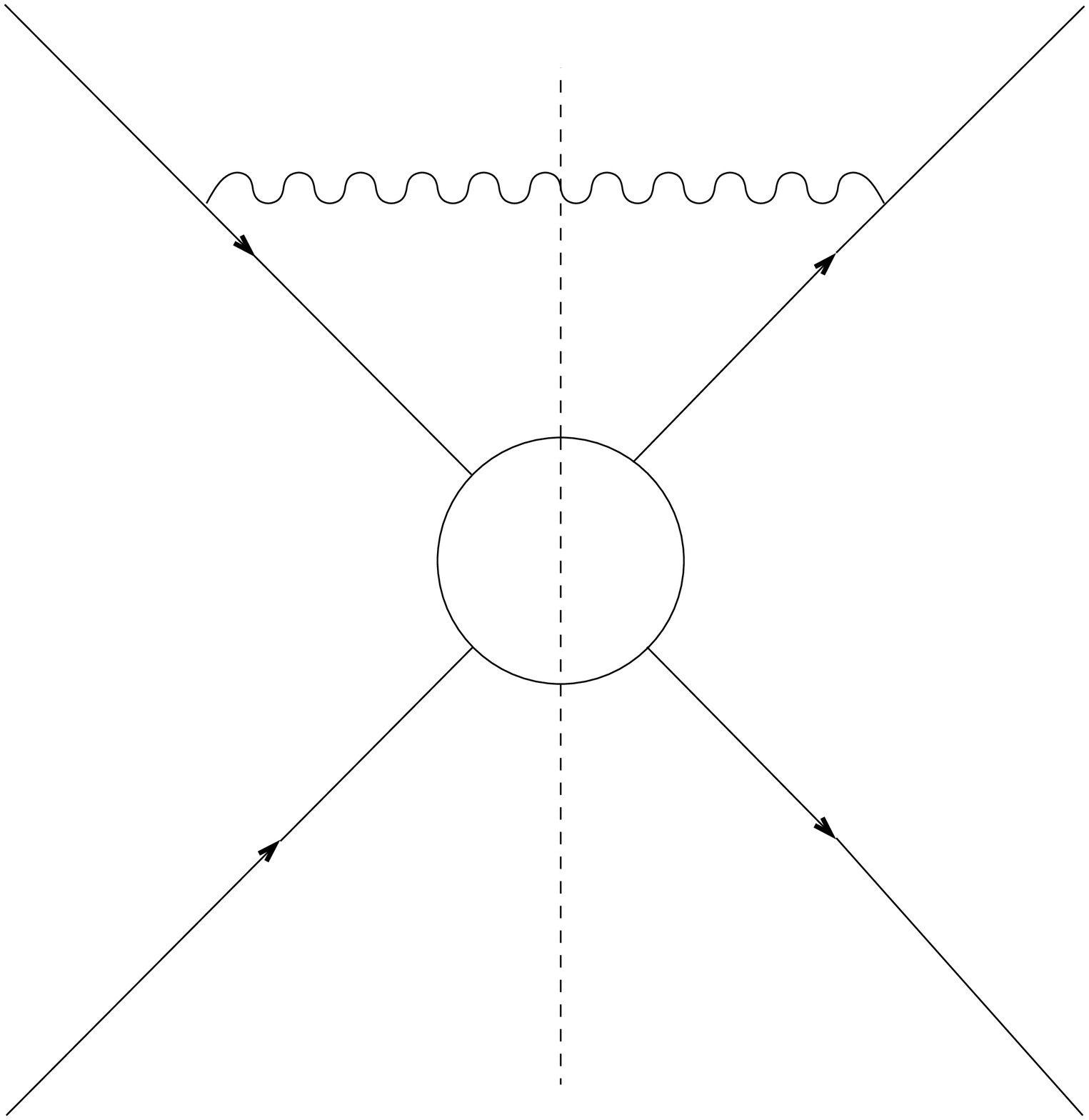,height=6cm}}
\put(3.6,6.5){$k,a$}
\put(.6,5.9){$t_1^a$}\put(5.5,5.9){$t_1'^a$}
\put(.7,6.8){$\alpha_1$}
\put(.7,0.9){$\alpha_2$}
\put(3.2,0.52){(a)}\put(12.2,.52){(b)}
\put(9.7,6){$t_1^a$}\put(10.7,5){$t_1^a$}
\put(9,6.8){1}\put(15.3,6.8){1'}
\put(9,0.9){2}\put(15.3,0.9){2'}
\put(6.4,6.8){$\beta_1$}\put(6.4,0.9){$\beta_2$}
\put(2.8,3.9){$S_H^\cro\,\, S_H$}
\put(9.3,1){\epsfig{file=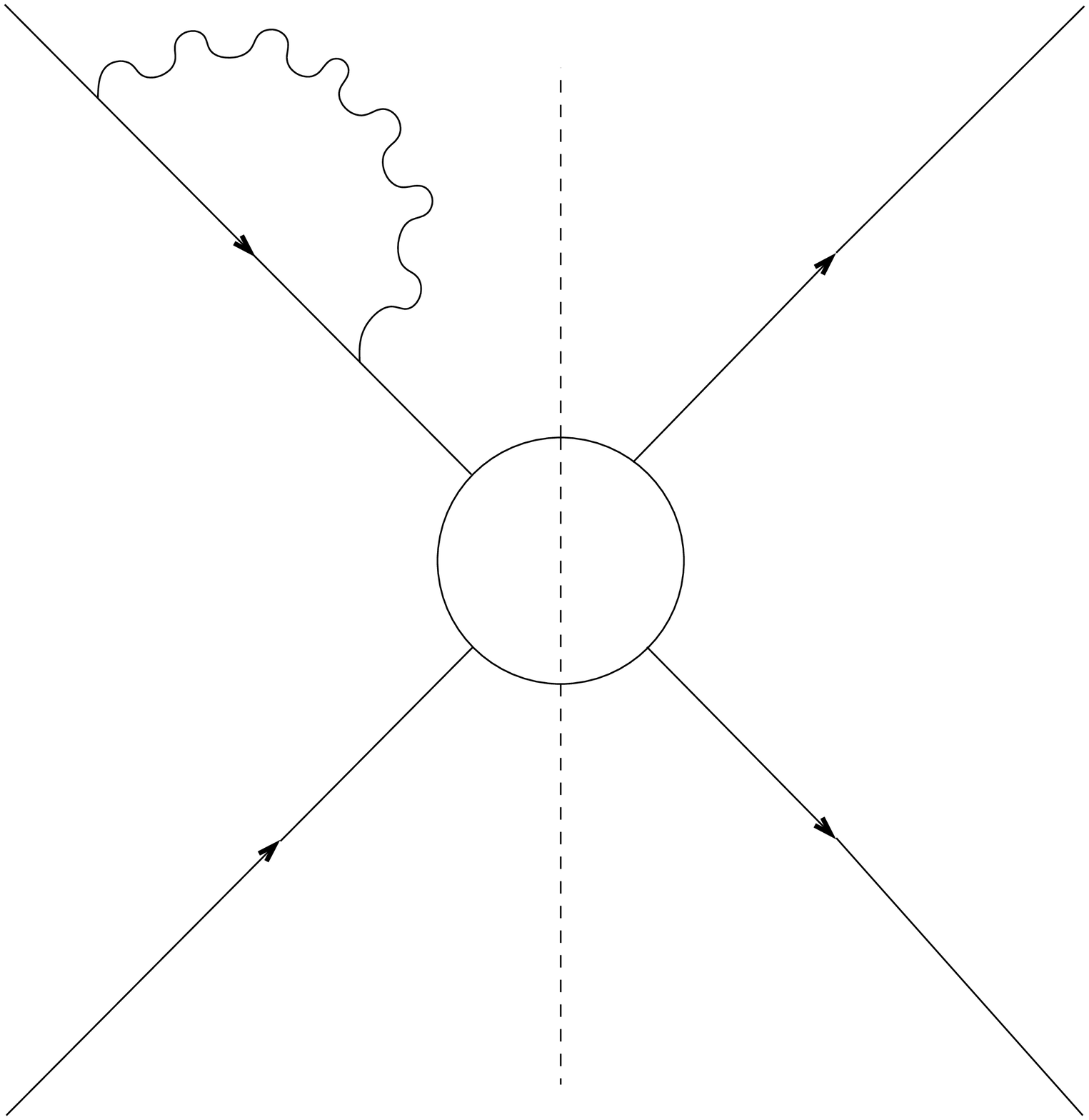,height=6cm}}
\put(11.8,3.9){$S_H^\cro\,\, S_H$}
\end{picture}
\caption{\label{1loopfig}
Unitarity diagrams for (b) virtual  and (a) real emission
 contributions to lowest order initial state interactions. 
Sum over gauge bosons a= $\gamma,Z,W$ is understood.}
\end{figure}

In isospin space, the hard cross section structure
is  defined by the so called
hard overlap matrix, describing the squared matrix element: 
$\bra \beta_1 \beta_2| S_H^\cro S_H|\alpha_1\alpha_2\ket\equiv 
\ov^H_{\beta_1\beta_2,\alpha_1\alpha_2}$ (see Fig. \ref{1loopfig}).
While for cross sections we always have
$\alpha_i=\beta_i$ ($\sigma_{\alpha\beta}\equiv 
\ov^H_{\alpha\beta,\alpha\beta}$), we leave open the possibility 
that $\alpha_i\neq\beta_i$ and see $O_H$ as an operator in isospin
space with four indices.

\begin{figure}[htb]\setlength{\unitlength}{1cm}
\begin{picture}(12,7.5)
\put(0.3,1){\epsfig{file=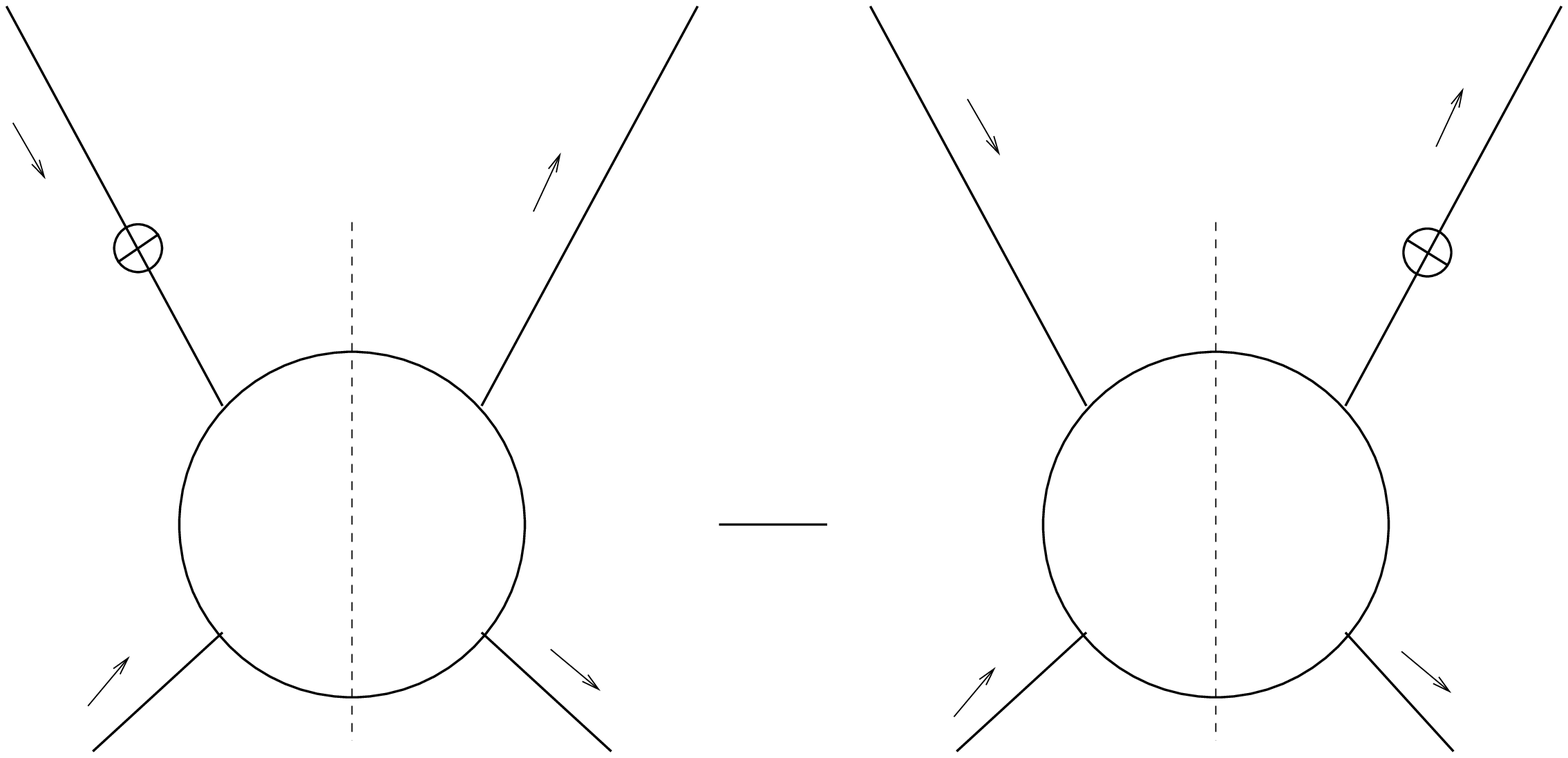,height=6cm}}
\put(0.1,5.7){$p_1$}\put(4.2,5.6){$p_1$}\put(1.7,5){$t_1^a$}
\put(0.7,1.7){$p_2$}\put(5,1.8){$p_2$}
\put(0.5,6.7){$\alpha_1$}\put(5,6.7){$\beta_1$}\put(2,4){$\delta$}
\put(7.5,6.7){$\alpha_1$}\put(12.7,6.7){$\beta_1$}\put(11.3,4.1){$\delta$}
\put(8.3,6.1){$p_1$}\put(11.4,6){$p_1$}\put(11,5){$t_1'^a$}
\put(7.7,1.7){$p_2$}\put(12,1.8){$p_2$}
\end{picture}
\caption{Operator insertions for the overlap matrix. 
For physical cross sections, $\alpha_1=\beta_1$, the contributions
from the neutral sector, that are proportional to 
$y_1-y'_1$ and $t^3_1-t'^3_1$, are identically zero.\label{figura2}}
\end{figure}

Since we work in the  SU(2)$\otimes$ U(1) symmetric limit, we can use 
isospin and hypercharge
conservation in the form (notice the minus sign, due to 
the choice of momentum flow in Fig. \ref{figura2})  
\be\label{2}
(y_1-y_1')=(y'_2-y_2)
\qquad\qquad
(t_1^a-t_1'^a)=(t_2'^a-t_2^a) \quad a=1,2,3
\ee
The isospin operator $t_1(t_1')$ acts on the $\alpha_1$ $(\beta_1)$
index, and so on.
Here comes the important step: we change basis 
and couple the indices $\alpha_1,\beta_1$ that refer to leg 1 
to form SU(2) multiplets; the same is done for leg 2. 
Then, eqn. (\ref{2}) can also be seen as the statement that
the operator $\ov^H$ commutes with the total isospin in the t-channel
$\Ttot\equiv\tvet_1-\tvet'_1$. From this, we can derive the form of
$\ov^H$:
\be\label{n3}
[\Ttot,\ov^H]=0\Rightarrow\bra t,t_3|\ov^H|t',t'_3\ket
=C_t\delta_{tt'}\delta_{t_3t_3'}\Rightarrow
\ov^H=\sum_t C_t {\cal P}_t\quad ;\quad
{\cal P}_t=\sum_{t_3}|t,t_3\ket\bra t,t_3|
\ee
so that $\ov^H$ has the form of a sum of isospin multiplets
projectors ${\cal P}_t$, with coefficients $C_t$. The sum runs only over
integers $t=0,1,2,...$ since we are coupling a particle and its own
antiparticle in the t-channel. The coefficients $C_t$ depend on the 
specific process considered, and contain all the dynamics; for 
some relevant examples, see \cite{3p}.

We now describe the dressing of the hard overlap matrix
$\ov^H$ to give the dressed one $\ov$ through soft weak 
bosons emission and
absorption with  the external (initial) line insertions of the
eikonal currents 
\be\label{correnti}
J_a^\mu=g[\frac{p_1^\mu}{kp_1}(t_1^a-t_1'^a)+
\frac{p_2^\mu}{kp_2}(t_2^a-t_2'^a)]
\qquad
J_0^\mu=g'[\frac{p_1^\mu}{kp_1}(y_1-y_1')+\frac{p_2^\mu}{kp_2}(y_2-y_2')]
\ee
Here a=1,2,3 is the SU(2) index, we work in the unbroken basis
$A_0=c_W \gamma+s_W Z,A_3=-s_W\gamma+c_WZ$, $g$ and $g'$ being the usual 
electroweak couplings with $\frac{s_W}{c_W}=\frac{g'}{g}$. 
 Notice that, as is discussed thoroughly in \cite{3p},
only initial state radiation needs to be included, 
since final state radiation automatically 
cancels between real and virtual corrections. 

Before proceeding, we need to discuss an important and subtle point. 
Eqn. (\ref{correnti}) includes, in the neutral sector, both the 
contributions of the massive $Z$ boson and the massless photon. 
However, $\gamma,Z$ insertions are  automatically cancelled out
when summing  real and virtual corrections;
this is true  already at the eikonal current 
insertion operator level, eqn. (\ref{correnti}). 
This happens because, when considering  cross sections,
the diagonal operators $y_1-y_1'$ and $t_1^3-t_1'^3$ 
give 0 contributions, since the quantum numbers of
leg 1 and leg 1' are obviously the same
(see Fig. \ref{figura2}). For the sake of
convenience, one can then reinsert the  vanishing $A_3$
contribution, 
ending up effectively with a SU(2) theory with 
infrared cutoff $M$. 
In other words, one can summarize like this: when
considering completely inclusive observables in the SU(2)$\otimes$U(1)
electroweak sector of the Standard Model at the DL level in 
the recovered symmetry limit, 
one  effectively needs to compute
only initial state radiation in the corresponding SU(2) theory with
gauge bosons $A^1_\mu, A^2_\mu, A^3_\mu$ and IR cutoff (effective mass)
$M\sim M_Z\sim M_W$.  Then, one ends up with the insertion operator:
\be\label{10}
I(k)\equiv
g^2\frac{2p_1p_2}{(kp_1)(kp_2)}(\tvet_1-\tvet_1')\cdot (\tvet_2-\tvet_2')
\qquad\quad (\tvet_1\cdot\tvet_2\equiv\sum_a t_1^at_2^a)
\ee
where the charge factor can be replaced, because of the 
conservation (\ref{2}), by
\be\label{n6}
(\tvet_1-\tvet_1')\cdot (\tvet_2-\tvet_2')=
-(\tvet_1-\tvet_1')^2=2\tvet_1\cdot\tvet_1'-\tvet_1^2-\tvet_1'^2
\ee
The latter expression provides the charge computation in the axial
gauge, because in this gauge  the $W$ emission and absorption takes place 
on the same leg,
for both virtual ($-2\tvet_1^2$) and real emission ($2\tvet_1\cdot\tvet_1'$)
contributions. We can see from Fig. \ref{1loopfig}
the reason for noncancellation between
virtual 
and real one loop corrections:
the crucial point is that while  $\gamma, Z$ emission does not change the
initial state, $W$ emission does. Then, in the $W$ case virtual corrections
and 
real corrections 
are of opposite sign but do not cancel completely.

From (\ref{n6}) we see that the eikonal insertion is described  by the 
Casimir operator
$\Ttot^2$.
The resummed expression for the overlap matrix is given by the 
following expression, involving the energy ordered ($w_1\ll w_2\ll
...w_n$ where $w_i$ are the soft bosons energies) product $P_w$:
\be\label{n9}
\ov(s)=P_w\{\int dk I(k)\}\ov^H=\exp[-\dl \Ttot^2]\ov^H
=\sum_t P_t C_t e^{-\dl t(t+1)}
\ee 
where we have defined the eikonal radiation factor for $W$ exchange:
\be
\dl=\frac{g^2}{2}\int_M^E\frac{d^3\kvet}{2w_k(2\pi)^3}
\frac{2p_1p_2}{(kp_1)(kp_2)}=\frac{\alpha_W}{4\pi}\log^2\frac{s}{M^2}
\qquad (\alpha_W=\frac{g^2}{4\pi})
\ee
and we have taken into account the fact that the SU(2) Casimir $\Ttot^2$ is 
diagonal on the projection operators $\Ttot^2 P_t=t(t+1)P_t$, turning
the
$w$-ordered exponential into a regular one.

Equation (\ref{n9}) is our main result, since it allows to know the
all order DL resummed
overlap matrix (and consequently the physical cross sections) at any 
energy $\sqrt{s}$, once the Born cross sections (or, equivalently,
the coefficients $C_t$) for a given hard process are known. 

It is clear from  (\ref{n9}) that 
the only component that survives at very high energies 
is the singlet one ($t=0$), that doesn't evolve with energy.
This component corresponds to the averaged cross section
$\bar{\sigma}=\sum_{i,j}\sigma_{ij}$ that is left unchanged by 
soft boson emission and gives the asymptotic value of all
cross sections.
On the other hand, all other components get bigger and bigger effects
as the isospin value gets higher, due to the coefficient
$-t(t+1)$ in exponents in  (\ref{n9}), and vanish asymptotically. 
This means, namely, that bigger effects can be present in the 
boson-boson scattering case, where two isospin 1 are composed,
with respect to the isospin $\frac{1}{2}$ case of fermion-fermion
scattering.  To be more specific, the spin 1 projector 
components $C_1$ get suppressed by  $(1-e^{-2\dl})\approx$
13 \% at 1 TeV, while the corresponding depression for 
the spin 2 components $C_2$  is $(1-e^{-6\dl})\approx$ 34 \%
at 1 TeV.

Let us now specify to the fermion-fermion case, already
analyzed in (\cite{3p}).
Lefthanded 
fermions belong to the fundamental, isospin 1/2, representation of
SU(2). The t-channel composition of two isospin 1/2 states gives rise
to isospin 0 and isospin 1 states. This means that only ${\cal P}_0$
and ${\cal P}_1$ are present in decomposition (\ref{n3}) of the Overlap
matrix. Using (\ref{n9}) and the explicit form of the projection
operators for this case, we obtain:
\be\label{n10}
\ov_{\beta_1\beta_2,\alpha_1\alpha_2}=C_0\delta_{\beta_1\alpha_1}
\delta_{\beta_2\alpha_2}+C_1
\tau^a_{\beta_2\alpha_2}\tau^a_{\beta_1\alpha_1}e^{-2\dl}
\qquad
\sigma_{\alpha_1\alpha_2}=\ov_{\alpha_1\alpha_2,\alpha_1\alpha_2}=C_0
+C_1\tau^3_{\alpha_2\alpha_2}\tau^3_{\alpha_1\alpha_1}e^{-2\dl}
\ee
from which we obtain ($\tau^a$ are the Pauli matrices):
\be
\sigma_{11}=\sigma_{22}=\frac{\sigma_{11}^H+\sigma_{12}^H}{2}
+\frac{\sigma_{11}^H-\sigma_{12}^H}{2}e^{-2\dl}
\qquad
\sigma_{12}=\sigma_{21}=\frac{\sigma_{11}^H+\sigma_{12}^H}{2}
-\frac{\sigma_{11}^H-\sigma_{12}^H}{2}e^{-2\dl}
\ee 
These results agree with the ones already obtained in \cite{3p}.

\section{Generalization to initial transverse bosons}

Initial partons in practice are either gauge bosons 
or fermions; thus the phenomenologically relevant cases are 
fermion-fermion scattering, already analyzed in the previous section, and
the boson-fermion and boson-boson cases that we consider here. 
The case of initial bosons deserves a special attention
due to the presence of  longitudinally polarized gauge bosons,  
that are peculiar since  they are sensitive to the symmetry breaking
Higgs sector even in the limit $\frac{M}{\sqrt{s}}\to 0$.
Leaving the case of longitudinal bosons to a more detailed forthcoming
analysis \cite{long}, we consider here only transverse gauge bosons, 
that we label with indices +,3,- for the triplet and 0 for the singlet. 
Note that since we rely only on SU(2) invariance, 
the index 0 can represent not only the $A_0$ gauge boson, 
but also  any SU(2) singlet like a gluon or a righthanded fermion.

Let us consider the case of fermion-boson scattering first.
Again, the composition of two isospin $\frac{1}{2}$
states on leg 1 produces only two projection operators; this time
however the explicit form of these operators is different from 
(\ref{n10}) due
to the presence of a isospin 1 state on leg 2:
\be\label{n12}
\ov_{\beta_1b_2,\alpha_1a_2}=C_0\delta_{\beta_1\alpha_1}
\delta_{b_2a_2}+C_1
T^a_{b_2a_2}\tau^a_{\beta_1\alpha_1}e^{-2\dl}
\qquad
\sigma_{\alpha_1 a_2}=\ov_{\alpha_1a_2,\alpha_1a_2}=C_0
+C_1T^3_{a_2a_2}\tau^3_{\alpha_1\alpha_1}e^{-2\dl}
\ee
where the generators in the adjoint representation are chosen to be:
\be
T^1=\frac{1}{\sqrt{2}}\left(\begin{array}{ccc}
0 & 1 & 0\\
1 & 0 & 1\\
0  & 1 & 0\end{array}\right)\qquad
T^2=\frac{1}{\sqrt{2}}\left(\begin{array}{ccc}
0 & -i & 0\\
i & 0 & -i\\
0  & i & 0\end{array}\right)\qquad
T^3=\left(\begin{array}{ccc}
1 & 0 & 0\\
0 & 0 & 0\\
0  & 0 & -1\end{array}\right)
\ee
and from which it is straightforward to obtain:
\be
\sigma_{1+}=\frac{\sigma_{1+}^H+\sigma_{1-}^H}{2}
+\frac{\sigma_{1+}^H-\sigma_{1-}^H}{2}e^{-2\dl}
\qquad
\sigma_{1-}=\frac{\sigma_{1+}^H+\sigma_{1-}^H}{2}
-\frac{\sigma_{1+}^H-\sigma_{1-}^H}{2}e^{-2\dl}
\ee 
\be
\sigma_{13}=\sigma_{23}=\frac{\sigma_{1+}+\sigma_{1-}}{2}
=\frac{\sigma^H_{1+}+\sigma^H_{1-}}{2}
\qquad\sigma_{2-}=\sigma_{1+}\qquad
\sigma_{2+}=\sigma_{1-}
\ee

In the case of boson-boson scattering 
the composition of two isospin 1 multiplets 
in the t-channel produces also a spin 2 projector, whose component 
is highly suppressed because of the factor $2(2+1)=6$. 
The general decomposition in this case is therefore:
\be
\ov_{b_1b_2,a_1a_2}=C_0\delta_{b_1a_1}
\delta_{b_2a_2}+C_1
T^a_{b_2a_2}T^a_{b_1a_1}e^{-2\dl}
+C_2[\{T^a,T^b\}_{b_1a_1}\{T^a,T^b\}_{b_2a_2}-\frac{16}{3}
\delta_{b_1a_1}\delta_{b_2a_2}]e^{-6\dl}
\ee
\be
\sigma_{a_1 a_2}=\ov_{a_1 a_2,a_1 a_2}=C_0
+C_1T^3_{a_2a_2}T^3_{a_1a_1}e^{-2\dl}+
C_2 \frac{2}{3}(\delta_{a_2 +}-2
\delta_{a_2 3} +\delta_{a_2 -})(3 (T^2_3)_{a_1 a_1}-2)e^{-6\dl} 
\ee
from which we derive the relations:
\be\label{eqn16}
\sigma_{--}=\sigma_{++}\qquad\sigma_{3-}=\sigma_{3+}\qquad\sigma_{33}=
\sigma_{++}+\sigma_{-+}-\sigma_{3+}
\ee
and, with $\sigma_{a}\equiv\sigma_{a+}$:

\begin{subequations}\label{eqn17}\begin{eqalignno}
\sigma_+&= \sigma_+^H(\frac{1}{3}+\frac{e^{-2\dl}}{2}+\frac{e^{-6\dl}}{6})
+\sigma_-^H(\frac{1}{3}-\frac{e^{-2\dl}}{2}+\frac{e^{-6\dl}}{6})
+\sigma_3^H(\frac{1}{3}-\frac{e^{-6\dl}}{3})
\\
\sigma_-&= \sigma_+^H(\frac{1}{3}-\frac{e^{-2\dl}}{2}+\frac{e^{-6\dl}}{6})
+\sigma_-^H(\frac{1}{3}+\frac{e^{-2\dl}}{2}+\frac{e^{-6\dl}}{6})
+\sigma_3^H(\frac{1}{3}-\frac{e^{-6\dl}}{3})
\\
\sigma_3 &=  \sigma_+^H(\frac{1-e^{-6\dl}}{3})
+\sigma_-^H(\frac{1-e^{-6\dl}}{3})+\sigma_3^H(\frac{1+2e^{-6\dl}}{3})
\end{eqalignno}\end{subequations}
We also give for completeness the expression for $\sigma_{33}$
that can be derived from the above equations:
\be
\sigma_{33}=
(\sigma_{-}^H+\sigma_{+}^H)\frac{1+2 e^{-6\dl}}{3}+
\sigma_{3}^H\frac{1-4 e^{-6\dl}}{3}
\ee

We now briefly discuss mixing in the weak bosons sector. 
The physical states in the neutral sector are linear combinations 
$(\gamma,Z)=M (A_0,A_3)$ where $M$ is the 2x2 matrix
$M_{11}=M_{22}=c_W, M_{12}=-M_{21}=s_W$. It is straightforward to 
obtain physical cross sections involving neutral gauge bosons on the external
legs. For instance, if $A,B$ are $\gamma, Z$ indices we have:
\be\label{eqn19}
\sigma_{AB}=\ov_{AB,AB}=\sum_{a_1,a_2,b_1,b_2} M_{Aa_1} M_{Ab_1}
 M_{Ba_2} M_{Bb_2}\ov_{b_1b_2,a_1a_2}
\ee
In this equation also overlap matrix elements that do not correspond
to physical cross sections appear; this is the case for $\ov_{00,33}$ 
for instance. 
Therefore we need to know also all overlap matrix elements involving 
the $A_0$ boson, that we label with an index 0, on the external legs. 
There are obviously 4 possible cases, with one, two three  or four 
$A_0$ bosons
on the external legs. Because of isospin invariance we can write:
\begin{subequations}
\be
\ov_{ij,k0}=\ov_{k0,ij}=C_1 T^{k}_{ij}e^{-2\dl}
\qquad
\ov_{ij,0k}=\ov_{0k,ij}=C_1 T^{k}_{ij}   e^{-2\dl}
\ee
\be
\ov_{ij,00}=\ov_{00,ij}=C_1 \delta_{ij}   e^{-2\dl}
\qquad
\ov_{0i,j0}=\ov_{i0,0j}=C_1 \delta_{ij}   e^{-2\dl}
\qquad
\ov_{i0,j0}=\ov_{0i,0j}=C_0 \delta_{ij}  
\ee
\be
\ov_{i0,00}=\ov_{0i,00}=\ov_{00,i0}=\ov_{00,0i}=0  
\ee
\be
\ov_{00,00}=C_0\equiv \sigma_{00}
\ee
\end{subequations}
where $i,j,k\neq 0$. From (\ref{eqn19}) we obtain the physical cross 
sections after accounting for mixing:
\begin{subequations}
\be
\sigma_{\gamma \gamma}= s_W^4 \sigma_{33}
+c^4_W  \sigma_{00}+2 s_W^2 c_W^2 
(O_{00,33}+O_{03,03}+O_{03,30})
\ee
\be
\sigma_{ZZ}= c_W^4 \sigma_{33}
+s^4_W  \sigma_{00}+2 s_W^2 c_W^2
(O_{00,33}+O_{03,03}+O_{03,30})
\ee
\be
\sigma_{\gamma Z}= \sigma_{Z \gamma}  = s_W^2 c_W^2( \sigma_{33}
+ \sigma_{00} )  
-2 s_W^2 c_W^2
O_{00,33}+ (c_W^4+s_W^4) O_{03,03}
-2 s_W^2 c_W^2 O_{03,30}
\ee
\be
\sigma_{ Z\delta}= 
 s_W^2  \sigma_{0\delta}
+
c_W^2  \sigma_{3\delta} +
2 s_W c_W  O_{3\delta,0\delta}\qquad(\delta=W^\pm,e,\nu,...)
\ee
\be
\sigma_{\gamma\delta}=  c_W^2 \sigma_{0\delta}  +
s_W^2  \sigma_{3\delta} - 2 s_W c_W  O_{3\delta,0\delta}
\qquad(\delta=W^\pm,e,\nu,...)
\ee
\end{subequations}

The formalism developed here is completely general. Given a specific process, 
all that one has to do is to find out the corresponding form
of the $C_i$ coefficients, given by the hard tree level cross sections. 
Besides the explicit examples 
in  \cite{3p} that refer to the fermion-fermion scattering case,
we give here an example with the purpose of seeing the general
formalism at work and of having an idea of the order of magnitude
of the effects that are expected. 
With this  in mind, let us consider the cross section for
initial transverse gauge bosons into two hadron jets, obtained
by summing over quark-antiquark final pairs. 
Two kind of diagrams contribute to the hard (tree level) cross section
for massless fermions: t,u-channel 
fermion exchange and s-channnel annihilation. In the limit
$g'\to 0$ it is easy to obtain the cross sections
\be\label{hard}
\frac{d\sigma^H_{++}}{d c_\theta}=0
\quad
\frac{d\sigma^H_{3+}}{d c_\theta}=
\frac{\pi \alpha_W^2 N_c N_f}{8 s}
\frac{c_\theta^2(1+c_\theta^2)}{s_\theta^2}
\quad
\frac{d\sigma^H_{33}}{d c_\theta}=
\frac{\pi \alpha_W^2 N_c N_f}{8 s}
\frac{(1+c_\theta^2)}{s_\theta^2}
\quad
\frac{d\sigma^H_{-+}}{d c_\theta}=
\frac{\pi \alpha_W^2 N_c N_f}{8 s}
\frac{(1+c_\theta^2)^2}{s_\theta^2}
\ee
where $\theta$ is the angle between the $W_-$ (or $A_3$) and the fermion.
The energy dependence is now obtained by inserting
the hard cross sections values (\ref{hard}) into the energy evolution
equations (\ref{eqn17}). Notice that the isospin relations (\ref{eqn16}) 
are satisfied, as expected,   for any energy value. 
The case of $\sigma_{++}$ is particularly interesting, since this cross 
section is zero  at the tree level. However, a $W^+$ can radiate 
a soft $W^+$ becoming an $A_3$ which has a sizeable tree level cross section.
Asymptotically,  $\sigma_{++}$ tends to the singlet value which is the 
cross sections average. At 1 TeV for instance
$\frac{\sigma_{++}}{\sigma_{3+}}\approx 12 \%$ so that $\sigma_{++}$, 
despite being 0 at tree level, reaches at the TeV scale a value comparable 
to the other cross sections. Analogous interesting effects 
can be found in the angular dependence since, for instance, 
$\sigma_{3+}^H$ is zero for $\theta=\frac{\pi}{2}$. 
The relative effects one finds 
for $\sigma^H_{33},\sigma^H_{-+}$ are
in the 12 - 20 \% range at the TeV scale. 
In conclusion, pretty large effects are expected
for boson fusion processes: a realistic calculation, though, has to
include the luminosity weights for the various initial states, and the
longitudinal contributions \cite{long} as well.



\end{document}